\documentclass[apjl, trackchanges]{emulateapj}
\usepackage{xcolor} 
\usepackage[backref,breaklinks,colorlinks,urlcolor=blue, citecolor={blue!50!black}]{hyperref}
\usepackage{bm}
\usepackage{calc}
\usepackage{amsmath}

\shorttitle{Gas content from thermal SZ effects}
\shortauthors{Lim et al.}
\begin{document}

\title{Gas contents of galaxy groups from thermal Sunyaev-Zel'dovich effects}

\author{S.H. Lim\altaffilmark{1},
H.J. Mo\altaffilmark{1,2},
Ran Li\altaffilmark{3},
Yue Liu\altaffilmark{1},
Yin-Zhe Ma\altaffilmark{4},
Huiyuan Wang\altaffilmark{5},
Xiaohu Yang\altaffilmark{6}}

\altaffiltext{1}
{Department of Astronomy, University of Massachusetts, Amherst MA 01003-9305; slim@astro.umass.edu}

\altaffiltext{2}
{Physics Department and Center for Astrophysics, Tsinghua University, Beijing 10084, China}

\altaffiltext{3}
{Key laboratory for Computational Astrophysics, National Astronomical 
Observatories, Chinese Academy of Sciences, Beijing, 100012, China}

\altaffiltext{4}
{School of Chemistry and Physics, University of KwaZulu-Natal,
Westville Campus, Private Bag X54001, Durban, 4000, South Africa; NAOC-–UKZN Computational Astrophysics Centre (NUCAC),
University of KwaZulu-Natal, Durban, 4000, South Africa}

\altaffiltext{5}
{Key Laboratory for Research in Galaxies and Cosmology,
Department of Astronomy, University of Science and Technology of China, 
Hefei, Anhui 230026, China; School of Astronomy and Space Science, University of 
Science and Technology of China, Hefei 230026, China}

\altaffiltext{6}
{Department of Astronomy, and Tsung-Dao Lee Institute,
Shanghai Jiao Tong University, Shanghai 200240, China; IFSA Collaborative Innovation Center, Shanghai Jiao Tong University, Shanghai 200240, China}

%%%%%
%%%HJM revised
%%%%
\begin{abstract}
A matched filter technique is applied to the {\it Planck} all-sky Compton $y$-parameter 
map to measure the thermal Sunyaev-Zel'dovich (tSZ) effect produced by galaxy groups 
of different halo masses selected from large redshift surveys in the low-$z$
Universe. Reliable halo mass estimates are available for all the groups,
which allows us to bin groups of similar halo masses to investigate 
how the tSZ effect depends on halo mass over a large mass range.
Filters are simultaneously matched for all groups to minimize projection 
effects. We find that the integrated $y$-parameter and the hot gas
content it implies are consistent with the predictions of the universal pressure profile model 
only for massive groups above $10^{14}\,{\rm M}_\odot$, but much 
lower than the model prediction for low-mass groups. The halo mass dependence 
found is in good agreement with the predictions of a set of simulations that 
include strong AGN feedback, but simulations including only supernova feedback 
significantly over predict the hot gas contents in galaxy groups. Our results suggest that hot gas 
in galaxy groups is either effectively ejected  
or in phases much below the virial temperatures
of the host halos.
\end{abstract}

\keywords{methods: statistical -- galaxies: formation -- galaxies: evolution -- galaxies: halos.}

%%%%%%% SECTION 1
%%HJM revised
\section[intro]{Introduction}
\label{sec_intro}

In the current paradigm of galaxy formation, galaxies are thought to 
form and evolve within dark matter halos \citep[see][for a review]{mo10}. 
During the formation of dark matter halos, the cosmic gas component 
first moves along with the dark matter, then gets shock-heated 
as the halos collapse, and eventually forms hot gaseous halos. 
In an adiabatic case, the resulting distribution of the hot gas 
component follows roughly that of dark matter, and the amount 
of the hot gas per dark matter is roughly a constant, about the 
universal baryon fraction of the universe. In reality, however, 
a myriad of other physical processes,  such as radiative cooling,
star formation, feedback from supernovae (SNe) and 
active galactic nuclei (AGN), etc, can change the hot gas 
content of the halos. Indeed, the hot gas fractions in low-mass halos
are found to be lower than the universal baryon fraction both  
in observations \citep[e.g.][]{david06, gastaldello07, pratt09, sun09}
and in numerical simulations \citep[e.g.][]{mccarthy10, battaglia13, lebrun14}. 
Even in massive systems, such as rich clusters of galaxies where the total 
hot gas is found to be closer to the universal value, the distribution of 
the hot gas is found to be different from that of dark matter 
\citep[e.g.][]{A10, battaglia12}. However, current observational 
results are still uncertain, particularly for low-mass systems, 
and many competing theoretical models have been proposed to 
describe the formation and structure of gaseous halos.
Clearly, an accurate determination of the hot gas content
in dark matter halos is crucial for understanding 
galaxy formation and evolution in a way complimentary to 
the information provided by stars and cold gas. 

The thermal Sunyaev-Zel'dovich effect \citep[tSZ hereafter;][]{sunyaev72} 
provides a promising avenue to probe the hot gas in halos.  
As the CMB photons pass through galaxy systems, such as clusters and groups 
of galaxies (collectively referred to as groups of galaxies),
they are scattered by the hot electrons by the inverse Compton process, 
producing a net energy gain in the photon gas and changing the 
CMB temperatures in the directions to the groups.  
Thus, studying the cross-correlation of the imprints of tSZ effect
on the CMB with galaxy groups allows one to probe the hot gas components 
in halos associated with galaxy systems. Compared to X-ray observations, 
the tSZ effect is less sensitive to the hot gas density, 
thus making it possible to explore the hot gas in the outskirts of 
halos and also in low-mass halos where the gas density is expected 
to be low.

However, extracting the tSZ signal reliably from CMB observations 
is not easy. First, the signal to be detected is usually comparable to 
or lower than the primary CMB, and there are also contaminations,
such as Galactic emissions, dust, and point sources. As a result, 
individual detection and analysis of the tSZ effect are currently 
only possible for rich clusters of galaxies \citep[e.g.][]{pcv}.
For low-mass groups, stacking of many systems is required to 
increase the signal-to-noise. Second, the beam sizes of current 
instruments are usually insufficient to resolve low-mass systems, 
so that assumptions about the spatial distribution of the hot gas are 
required. Finally, the signal from low-mass systems can be contaminated
by the projections of larger halos along the same line-of-sights, 
and such contamination is not straightforward to eliminate.  

Recently, \citet{pcxi} used the all-sky {\it Planck} Compton parameter 
map and a sample of locally brightest galaxies as tracers of dark matter 
halos to investigate the tSZ effects produced by galaxy systems
with halo masses down to $\sim 4\times10^{12}\,{\rm M}_{\odot}$. 
Remarkably, their results show that the universal pressure profile (UPP) model, 
in which the hot gas fraction relative to halo mass is independent 
of halo mass, matches their data well. This finding is in conflict with the 
results obtained from X-ray observations and hydrodynamic simulations 
where a much lower fraction is found for hot gas in low-mass systems. 
Using a hydrodynamic simulation, \citet{lebrun15} found that 
the universal pressure profile of \citet{A10} adopted by \citet{pcxi}
in their matched filter method may lead to over-estimations of the 
integrated tSZ signal for low-mass systems. However, \citet{greco15} 
showed that adopting another popular pressure profile of 
\citet{battaglia12}, instead of that of \citet{A10}, leads 
to differences that are well within the observational uncertainties, 
and so the high gas fraction found by \citet{pcxi} cannot be explained
by the adopted profile. \citet{Ma15} cross-correlated the {\it Planck} tSZ map with gravitational lensing map from CFHTLenS survey and found that the prediction of UPP model is $~20\%$ higher than the data. \citet{vikram17} cross-correlated the {\it Planck} 
tSZ map with the group catalog of \citet{yang07} and found that the 
two-halo terms dominate the tSZ signal for systems of 
$M_{200}\leq10^{13}\,h^{-1}{\rm M_\odot}$, indicating that projection 
effect is an important issue.  

In this paper, we extract the tSZ signal from the {\it Planck} all-sky 
observation for galaxy systems of different halo masses, 
using the group catalog of \citet{lim17}. The catalog is constructed 
for four large redshift surveys with the use of the halo-based group finder of 
\citet{yang05, yang07}. This provides the largest sample of galaxy 
groups in the low-$z$ universe to study the tSZ effects over a large 
range of galaxy systems. In particular, reliable halo mass estimates
are provided for all groups, so that we can bin groups of similar 
masses to investigate how the tSZ effect depends on halo mass.    
We employ the matched filter technique \citep{haehnelt96, herranz02, 
melin05, melin06} to extract the tSZ signal from the {\it Planck} map. 
In particular, we simultaneously match the filters to all 
galaxy systems in the catalog, so that projection effects 
produced by halos along the line-of-sights are properly 
taken care of. 

The outline of this paper is as followings. We describe the observational 
data used in our analysis in Section~\ref{sec_data}, and our method to extract
the tSZ signal in Section~\ref{sec_method}. We present our main results as well 
as comparisons with results from earlier studies and from 
numerical simulations in Section~\ref{sec_result}. Finally, 
we summarize and conclude in Section~\ref{sec_sum}. 

We adopt the cosmological parameters from the {\it Planck} observation \citep{pcxiii},
$\{\Omega_{\rm m}, \ \Omega_{\Lambda},\ h, \ \sigma_8\}$=$\{0.308, 0.692, 0.678, 0.831\}$
throughout this paper unless specified otherwise. 

%%%%%%% SECTION 2
%% HJM revised
\section[data]{Observational Data}
\label{sec_data}
\subsection{The {\it Planck} $y$-map} 
\label{ssec_Planck} 

The {\it Planck}~\citep{tauber10, pci}, a space mission 
to measure the CMB anisotropy, is an all-sky observation in nine 
frequency bands ranging from $30$ to $857$ GHz, 
with angular resolutions from $31\arcmin$ to $5\arcmin$. 
For our analysis of the thermal Sunyaev-Zel'dovich (tSZ)
effects, we use the {\it Planck} NILC \citep[Needlet Independent Linear
Combination;][]{remazeilles11} all-sky tSZ Compton parameter
map \citep{pcxxii}, also referred to as the NILC $y$-map, 
which is part of the publicly released {\it Planck} 2015 
data\footnote{\url{https://pla.esac.esa.int}}. The map
is constructed from the full mission data set, using 
a combination of different frequency maps to remove the 
primary CMB fluctuations and to minimize contamination
from foreground sources. For more details of the $y$-map 
construction, the readers are referred to the original 
paper cited above. To limit the Galactic foreground contamination,  
which is mainly due to thermal dust emissions, we mask the brightest 
40\% of the sky by applying the corresponding mask provided in 
the {\it Planck} 2015 data release. For contamination from extra-galactic
sources, such as radio and infrared galaxies, we apply the 
mask provided in the same data release for point sources.

\subsection{Galaxy groups} 
\label{ssec_group} 

In order to determine the tSZ signals from halos associated 
with different galaxy systems, we need a well-defined group catalog 
that provides reliable information for both the positions 
and halo masses of the galaxy systems in the universe. 
Furthermore, since the tSZ signals are typically weak for individual 
groups, and since it is necessary to stack many systems to increase 
the signal to noise ratio, a well-defined group
catalog is also needed to interpret the stacking results. 
In this paper we use the group catalogs given in \citet{lim17}, 
which uses four redshift catalogs of galaxies (2MRS, 6dF, SDSS, and 2dF)
to achieve an almost all-sky ($91\%$) coverage and the best depth reachable 
by these galaxy catalogs in each region of the sky. 
Groups are identified with the adaptive halo-based group finder of 
\citet{yang05, yang07} with some modifications (see \citet{lim17} for
the detail). Tests with realistic mock galaxy catalogs 
show that the halo masses assigned by the group finder match
well the true masses,  with typical scatter of $0.2-0.3\,{\rm dex}$.  
The catalogs provide two different halo mass estimates based either 
on the luminosities and stellar masses of member galaxies, and we 
use the masses based on the stellar masses. We combine   
2MRS, 6dF, and SDSS to construct our sample of groups 
with $\log M_{500}/{\rm M_\odot}\geq 12$ for the tSZ
analysis. For sky regions covered by more than one catalog, 
the preference is given in the order of SDSS, 6dF, and 2MRS. 
The sample contains a total of $471,696$ galaxy systems (groups), 
of which $3,851$ have $\log M_{500}/{\rm M_\odot}\geq14$, 
$112,494$ have $13\leq\log M_{500}/{\rm M_\odot}\leq14$, 
and $240,747$ have $12\leq\log M_{500}/{\rm M_\odot}\leq13$.
Following conventions in previous SZ effect analyses, 
we define a halo by a radius $R_{500}$, within which the mean 
density is $500$ times the critical density at the redshift in question. 
The mass, $M_{500}$, used above is the halo mass within $R_{500}$. 
The halo masses and radii provided in the group catalogs are 
$M_{200}$ and $R_{200}$, respectively. To convert these quantities 
to the corresponding $M_{500}$ and $R_{500}$, we assume NFW profiles 
\citep{navarro97} and concentration parameters as given 
by \citet{neto07}. 

%%%%%%% SECTION 3
%% SHL revised 
\section[method]{Method and analysis}
\label{sec_method}

\subsection{The matched filter technique} 
\label{ssec_mf} 

Detecting the SZ signals physically related with a galaxy system is 
not trivial, as other effects, such as the primary 
CMB anisotropies, Galactic foreground, and other sources, can all contaminate 
the signals we want to obtain (see Section~\ref{sec_result}). 
Using a simple aperture photometry to extract the signals may thus lead to 
large uncertainties in the extracted signals \citep[see e.g.][]{melin06}.
To limit source confusions and background contamination, 
we employ the matched filter (MF) technique, first proposed for SZ 
analyses by \citet{haehnelt96}, which is designed to maximize the signal-to-noise
for a SZ source by imposing prior knowledge of the signals given the noise
power spectra. For the case considered here, this means to optimally 
extract the tSZ signals from groups of galaxies, under the constraint 
of the power spectrum of the noise of the {\it Planck} maps.
In practise, we closely follow \citet{melin05, melin06}, who presented 
an extended and general formalism to extract signals from SZ surveys
using the multi-filtering technique of \citet{herranz02}.
Such a MF technique has been applied in many recent analyses of 
the SZ effects in different surveys \citep[e.g.][]{pcv, pcxi, pcliii, li14, lebrun15}. 

%%%%%%%%%%% figure1
\begin{figure}
\includegraphics[width=1.\linewidth]{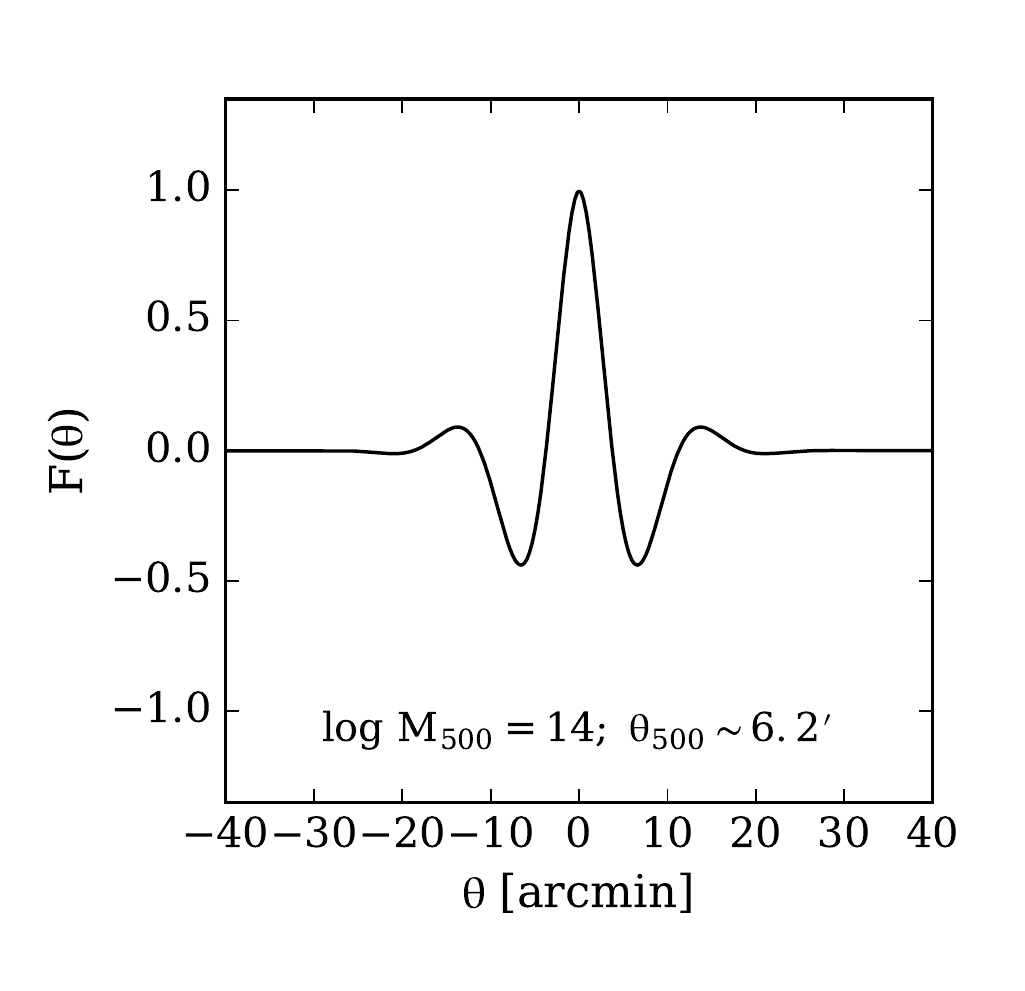}
\caption{An example of the matched filter constructed for the tSZ analysis. 
Here a universal profile of \citet{A10} is adopted as a spatial filter 
for a group of $\log M_{500}/{\rm M_\odot}=14$ and an angular 
radius $\theta_{500}\sim 6.2\arcmin$.}
\label{fig_filter}
\end{figure}

In the MF approach, the Fourier transform of the filter that maximizes
the signal-to-noise is given by: 
%%%%%%%%%%%%%%%%%%%%%%%%%%%%%%%%%%
\begin{eqnarray}
\hat{F}(\boldsymbol{k}) = \Big[\int \frac{|\hat{\tau}(\boldsymbol{k'})\hat{B}(\boldsymbol{k'})|^2}
{P(k')} 
\frac{{\rm d}^2k'}{(2\pi)^2}  \Big]^{-1} \frac{\hat{\tau}(\boldsymbol{k}) \hat{B}(\boldsymbol{k})}
{P(k)}
\end{eqnarray}
%%%%%%%%%%%%%%%%%%%%%%%%%%%%%%%%%%%
where $\hat{\tau}(\boldsymbol{k})$ is the Fourier transform of the assumed 
spatial profile of groups, $\hat{B}(\boldsymbol{k})$ is the Fourier transform
of a Gaussian beam function that mimics the 
convolution in {\it Planck} observation, and $P(k)$ is the noise power spectra. As the 
NILC $y$-map used here is already cleaned of the primary CMB anisotropies, 
$P(k)=P_{\rm noise}$, where $P_{\rm noise}$ is the power spectrum of the 
{\it Planck} noise map, as provided in the data release. 
The choice of the spatial filter function is
not straightforward, and it can affect  the integrated signals extracted. 
Indeed, using hydrodynamic simulations, \citet{lebrun15} found that 
the extracted tSZ signals can change significantly depending on 
the filter shape adopted. In our analysis, 
we adopt the universal pressure profile (UPP) given in
\citet[][hereafter A10]{A10}, the form of which can 
be written as
\begin{equation}
    P(r) =A [E(z)]^{8/3}{\cal P}(r/R_{500}, M_{500})\,,
\end{equation}
where $E(z)\equiv H(z)/H_0$, ${\cal P}$ specifies the 
shape of the profile, and $A$ is an overall amplitude
(see A10 for details). This profile was derived from a combination of 
X-ray observations of XMM-Newton REXCESS cluster sample \citep{bohringer07} 
at $r\leq R_{500}$ and hydrodynamic simulations at larger radii.
As a test, we have also used the spatial filter adopted in 
\citet{lebrun15} but did not find any significant changes 
in our results. 

Figure~\ref{fig_filter} shows an example of the constructed filters,  
which assumes a group with  $\log (M_{500}/{\rm M}_\odot)=14$ and 
an angular radius $\theta_{500}\sim 6.2^\prime$, and the universal 
pressure profile given above. 

\subsection{Extracting the tSZ signal}
\label{ssec_tSZ}

Theoretically, the tSZ signal is characterized by a Compton $y$-parameter, 
%%%%%%%%%%%%%%%%%%%%%%%%%%%%%%%%%%
\begin{eqnarray}
y\equiv \frac{\sigma_{\rm T}}{m_{\rm e}c^2} \int{P_{\rm e}\,{\rm d}l}, 
\end{eqnarray}
%%%%%%%%%%%%%%%%%%%%%%%%%%%%%%%%%%%
where $\sigma_{\rm T}$ is the Thompson cross-section, $m_{\rm e}$ the rest-mass of 
electron, $c$ the speed of light, $P_{\rm e}=n_{\rm e}k_{\rm B}T_{\rm e}$ the electron pressure, 
and the integration is over the line-of-sight to the observer. 

The filters described above are then put at the group centers and `matched' 
to the $y$-map to yield an estimate of the tSZ flux within $R_{500}$,
$Y_{500}$ defined by 
%%%%%%%%%%%%%%%%%%%%%%%%%%%%%%%%%%
\begin{eqnarray}
d_{\rm A}(z)^2 Y_{500} = \frac{\sigma_{\rm T}}{m_{\rm e}c^2} \int_{R_{500}}{P_{\rm e}\,{\rm d}V}, 
\end{eqnarray}
%%%%%%%%%%%%%%%%%%%%%%%%%%%%%%%%%%%
where $d_{\rm A}(z)$ is the angular diameter distance to a group at redshift $z$. 
Since $Y_{500}$ depends mainly on halo mass at a given $z$  
and evolves with $z$ as $E^{2/3}(z)$,
at a fixed halo mass, it is useful to define a new quantity,   
%%%%%%%%%%%%%%%%%%%%%%%%%%%%%%%%%%
\begin{eqnarray}
\tilde{Y}_{500}\equiv Y_{500}E^{-2/3}(z)
\Big({d_{\rm A}(z)\over 500 {\rm Mpc}}\Big)^2, 
\end{eqnarray}
%%%%%%%%%%%%%%%%%%%%%%%%%%%%%%%%%%%
which is expected to be a function of only halo mass scaled to $z=0$, if 
the intrinsic tSZ flux is indeed only a function of mass.  

To extract the SZ signals associated with galaxy groups 
from the observed $y$-map, a matched filter is put at 
each of all the groups in our group sample according to its 
halo mass and redshift. We then tune simultaneously 
the amplitudes of the filters for individual $M_{500}$ bins
as listed in Table~\ref{tab_Y500}, assuming the amplitudes 
for all the groups in a given $M_{500}$ bin to be the same.  
The overall best match between the matched filters and the observed $y$-map 
is sought on the basis of the sum of the $\chi^2$ over all the pixels 
covered by the filters. The simultaneous matching of individual groups 
allows us to take into account the line-of-sight contributions 
from other halos, which is important, particularly for low-mass 
systems (see next section). For most of our analyses, 
we truncate the filter at $3\,\theta_{500}$ and only use
fluxes within it to estimate the integrated flux,   
where $\theta_{500}=R_{500}/d_{\rm A}$. Note that $R_{500}
\sim 0.5\,R_{200}$. As shown later in Section~\ref{sec_result}, 
truncation of the filters at $10\,\theta_{500}$ leads to 
little change in our results. Finally, the mean flux within $R_{500}$
for groups in a given mass bin is estimated from the  
assumed spatial profile together with the amplitude 
obtained from the best match. The fluxes within $R_{500}$
for individual groups in a $M_{500}$ bin are also estimated  
by fixing the amplitudes of the matched filters of other 
mass bins to their best fitting values, while tuning 
the amplitudes of the filters for individual groups
in the $M_{500}$ bin in question to achieve the best 
match. 

%%%%%%% SECTION 4
%% Revised by HJM
\section[result]{Results}
\label{sec_result}

\subsection{The $M_{500}$-$\tilde{Y}_{500}$ relation and the hot gas content} 
\label{ssec_Y500}

%%%%%%%%%%% figure2
\begin{figure}
\includegraphics[width=1.\linewidth]{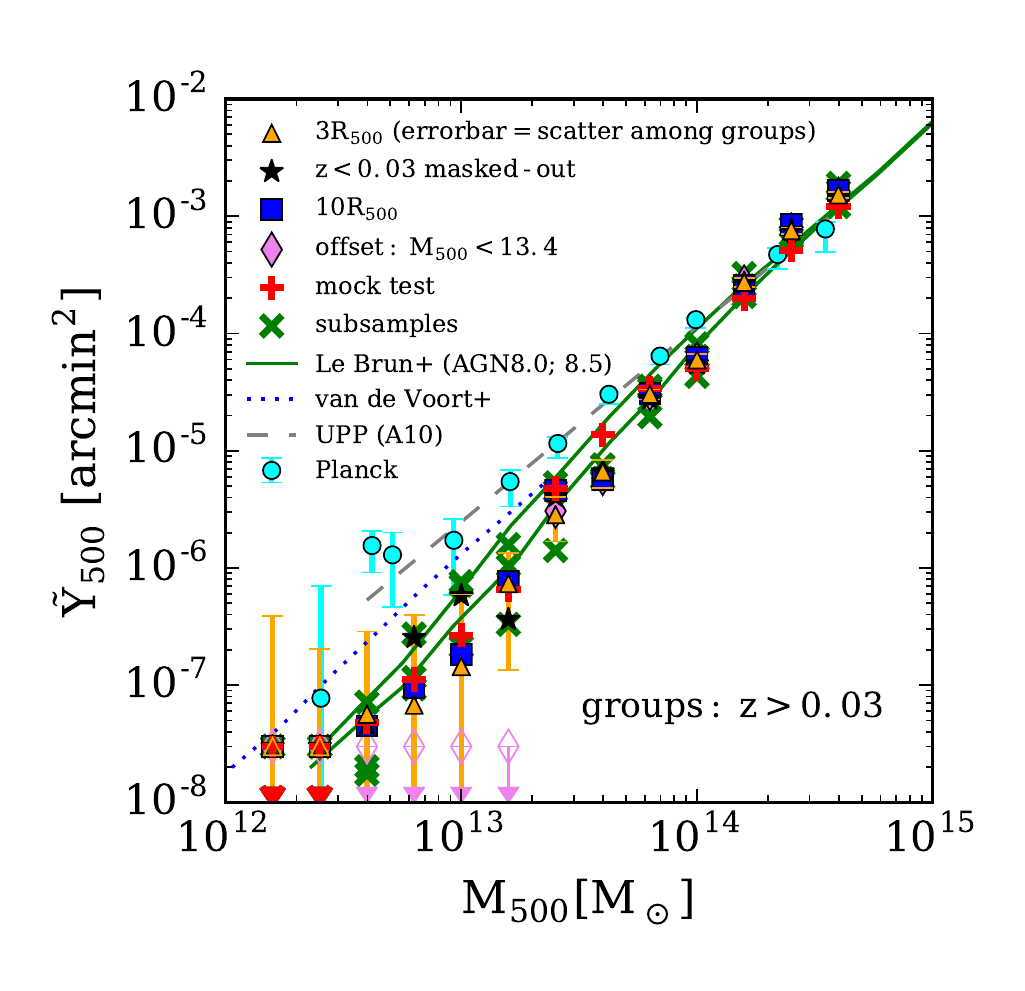}
\caption{The $M_{500}$-$\tilde{Y}_{500}$ relations obtained by applying the 
matched filters, with shapes following that of \citet{A10} (A10) 
truncated at $3R_{500}$ (triangles) and $10R_{500}$ (squares), 
respectively, and by masking out the pixels covered by groups at 
$z<0.03$ (stars). The case where the filters for groups
with $\log(M_{500}/{\rm M}_\odot)< 13.4$ are offset by 
$3\theta_{500}$ is shown by diamonds. The red crosses plot
the difference between the results from the co-added map and the
added component (see text for details of this test). 
The green crosses show the results obtained from three 
independent sub-samples of total sample. The results are 
compared with those from \citet{pcxi} (cyan dot), the universal 
pressure profile of A10 (UPP; dashed), \citet{lebrun15} (the two solid 
lines, with the upper one for their AGN8.0 and the lower one for AGN8.5),
and \citet{vdV16} (dotted line). The error bars indicate the scatters 
among the signals from individual systems in each mass bin. 
The unfilled symbols with downward arrows are 
used for cases where the tSZ fluxes are negative.}
\label{fig_Y500}
\end{figure}

%%%%%%%%%%% figure3
\begin{figure}
\includegraphics[width=1.\linewidth]{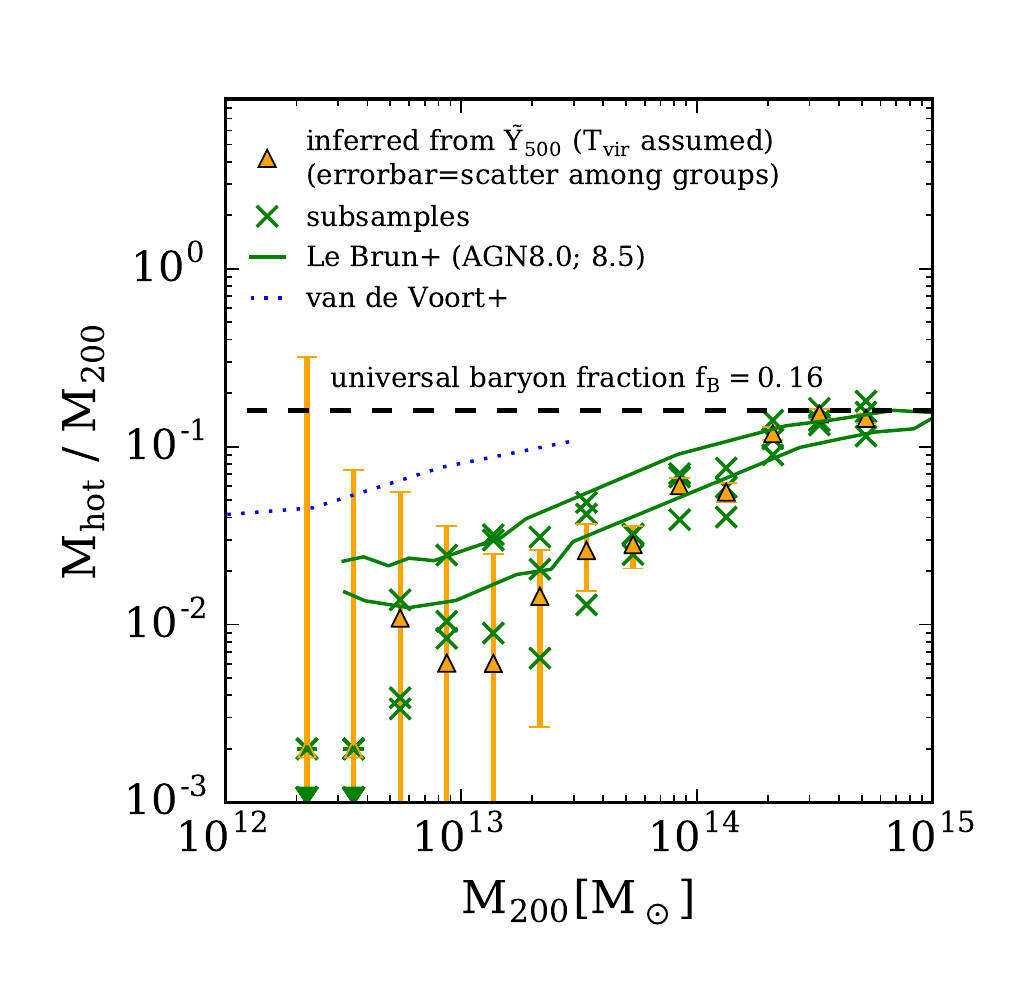}
\caption{Hot gas mass fraction with respect to halo mass within 
$r_{200}$ as a function of halo mass 
inferred from $\tilde{Y}_{500}$ by assuming the virial temperature, compared with
that from \citet{lebrun15} (solid; the upper for AGN8.0 and the lower
for AGN8.5) and \citet{vdV16} (dotted), 
which are based on hydrodynamic simulations. The gas content is lower than the 
universal baryon fraction of $f_B=0.16$ (dashed) 
in low-mass systems by a factor of up to $\sim 10$. Note that the gas mass ratio
estimated is inversely proportional to a temperature assumed. 
The error bars indicate the scatters of the signals among individual 
systems in each mass bin. The crosses show the results obtained 
from three independent sub-samples.  
The unfilled symbols are used for those with the tSZ flux below zero.}
\label{fig_gas_frac}
\end{figure}

\begin{deluxetable}{cccc}%[b!]
\tablecaption{The $M_{500}$-$\tilde{Y}_{500}$ relation\tablenotemark{a} 
\label{tab_Y500}}
%\tablenum{1}
%\tablewidth{0pt}
\tablehead{
\colhead{$\log M_{500}/{\rm M_\odot}$} &
\colhead{$\tilde{Y}_{500}$} &
\colhead{Scatter\tablenotemark{b}} & 
\colhead{No. of systems}\\
\colhead{} & 
\colhead{$[10^{-6}\ {\rm arcm}^2]$} &
\colhead{$[10^{-6}\ {\rm arcm}^2]$} &
\colhead{}
}
\startdata
$12.3$ & $-0.0600$ & $0.360$ & $40,689$ \\
$12.5$ & $-0.0480$ & $0.174$ & $41,848$  \\
$12.7$ & $0.0564$ & $0.231$ & $40,521$  \\
$12.9$ & $0.0675$ & $0.330$ & $37,344$  \\
$13.1$ & $0.144$ & $0.450$ & $32,063$  \\
$13.3$ & $0.735$ & $0.600$ & $25,744$  \\
$13.5$ & $2.85$ & $1.17$ & $19,020$  \\
$13.7$ & $6.60$ & $1.77$ & $12,500$  \\
$13.9$ & $30.3$ & $3.00$ & $6,203$  \\
$14.1$ & $59.7$ & $7.50$ & $2,163$  \\
$14.3$ & $273$ & $22.8$ & $484$  \\
$14.5$ & $756$ & $45.3$ & $195$  \\
$14.7$ & $1520$ & $83.7$ & $71$  
\enddata
\tablenotetext{a}{These data are presented in Fig.~\ref{fig_Y500} 
by triangles.}
\tablenotetext{b}{These are the scatters among individual systems in 
each mass bin.}
%\tablecomments{dd}
\end{deluxetable}

Figure~\ref{fig_Y500} shows our main result for the tSZ flux-halo 
mass relation. The values obtained from our fiducial sample 
are also given in Table~\ref{tab_Y500} for reference. 
In all our analyses, we use only groups at $z>0.03$, 
to avoid the domination by a small number of nearby groups 
each covering a large number of pixels. The error-bars 
shown are each the $1\sigma$ scatter among individual 
groups in the corresponding mass bin. For comparison,  
we also divide the total sample into three independent 
sub-samples that have about the same amount of sky coverage, 
and apply the MF method to them separately. The results are 
shown by the green crosses in the figure. As one can see, 
the scatter among these sub-samples is comparable to 
that among individual groups, indicating that the uncertainties 
are dominated by systematic effects. 

Comparing the result obtained from the fiducial case, where 
the matched filters are truncated at $3R_{500}$ with that 
obtained by using a larger truncation radius, $10R_{500}$, 
shows that $3R_{300}$ is sufficient to cover the signals
from individual groups.

Because the hot gas halos of groups are extended and 
the observational beam size is relatively large,  projection 
effects may contaminate the signals, particularly for small 
groups, even though our simultaneous matching of filters is 
supposed to eliminate such effects. We thus carry out a number 
of further tests to examine any possible residuals due to
projection effects. In our first test, we remove all pixels 
covered by the groups at $z<0.03$, and the results are shown 
by stars in Fig.~\ref{fig_Y500}. As one can see, the signals 
for low-mass systems are changed relative to the fiducial 
case, suggesting that our results for low mass groups 
may still be affected by projection effects. In the second 
test, the filters for groups below a halo mass limit 
are shifted by a given amount with respect to the group 
centers. Thus, if the signals extracted for these groups 
were not associated with them but produced by larger 
structures, such a shift would not change the signals
obtained for these groups. As an example, 
the diamonds in Fig.~\ref{fig_Y500}
show the result in which filters for groups with 
$\log M_{500}/{\rm M}_\odot<13.4$ are shifted randomly 
by $3R_{500}$, while the filters for more massive groups
are still located at the group centers. As one can see, 
while no significant change is seen for groups with
$\log M_{500}/{\rm M}_\odot>13.4$, 
the signals for the lower mass groups are reduced.
This suggests that the signals detected in the matched 
filters are associated with these low-mass groups. 
Finally, we make another test by adding to each group an artificial 
$y$-parameter profile which is given by the observed mean profile 
corresponding to its mass and redshift. The matched filter
technique is then applied to the sum of this artificial map
with the original map. The original signals are 
well recovered by the differences between the results 
obtained from the co-added map and the added component, 
as shown by the crosses in Fig.~\ref{fig_Y500}, 
demonstrating that our method can extract the signals 
we put in reliably. 

Based on the test results presented above, we conclude that 
the results for groups with masses above $10^{13.5}{\rm M}_\odot$ 
are stable. For groups of lower masses, however, significant
variations are still present from sample to sample.

Assuming virial temperatures, we estimate the hot gas 
contents of galaxy groups within $R_{200}$ from the integrated fluxes of 
$\tilde{Y}_{500}$. Here the NFW profile and the hot gas profile 
of A10 are used to convert quantities to the corresponding 
ones within $R_{200}$, and the virial temperature is defined as
%%%%%%%%%%%%%%%%%%%%%%%%%%%%%%%%%%
\begin{eqnarray}
T_{\rm vir}=\frac{\mu m_{\rm p} GM_{200}}{2k_{\rm B} R_{200}}
\end{eqnarray}
%%%%%%%%%%%%%%%%%%%%%%%%%%%%%%%%%%%
where $\mu$ is the mean molecular weight, $m_{\rm p}$ the proton mass, 
and $k_{\rm B}$ the Boltzmann constant. 
The results obtained from the fiducial sample and the 
three sub-samples are shown in Fig.~\ref{fig_gas_frac}. 
Here we see that the inferred hot gas 
contents of low-mass groups within $R_{200}$ are 
lower than the universal baryon fraction, 
shown by the horizontal line, by a factor of $\sim 10$. 
Even for groups with 
$M_{200}\sim 10^{14}{\rm M}_\odot$, the hot gas fraction is 
only about a half; only in the most massive groups (clusters) is
the fraction close to unity. Such low hot gas contents in 
low-mass groups have important implications for theories of 
galaxy formation, as to be discussed in the following.

\subsection{Comparisons with earlier results and theoretical models}

\citet{pcxi} (PCXI hereafter) used the same {\it Planck} data and a 
similar matched filter approach
to extract the tSZ signals around locally brightest galaxies (LBGs)
selected from the SDSS survey. An isolation criterion is adopted so that 
each LBG is the dominating one (in terms of luminosity) in its 
neighborhood, probably representing the central galaxy of a halo. 
Based on the mean relation between the stellar masses of central galaxies 
and the halo masses obtained from the semi-analytic galaxy formation 
model of \citet{guo13}, a halo mass is assigned to each of the LBGs. 
The $\tilde{Y}_{500}$-$M_{500}$ relation obtained by PCXI
is plotted in Fig.~\ref{fig_Y500} as circles, and matches well 
the expectation of the UPP model of A10, shown 
by the dashed line. As one can see, our results are in good agreement 
with that of PCXI only for massive groups with 
$M_{500}>10^{14}{\rm M}_\odot$, but the amplitudes we obtain 
for groups of lower masses are much lower. Indeed, our 
$\tilde{Y}_{500}$-$M_{500}$ relation is very different from that 
given by the UPP model. We suspect that there are two 
factors that may cause the difference between our and PCXI
results. First, we simultaneously match all groups in our 
sample, which takes into account the projection effects
by larger halos along the line-of-sights of low-mass groups, 
while PCXI matches individual filters separately. In a test 
where we first subtracted the local flat backgrounds averaged over 
annulus between $[2R_{200}, 3R_{200}]$ around each group, and  
then matched individual filters and stacked the signals for groups 
of similar masses, we found that we can roughly recover the results 
of PCXI for low-mass halos, despite of the differences in other 
details between our method and theirs. This indicates that the 
contamination by other groups is not flat, and that it is important to match 
the filters to all groups simultaneously in order to correct 
for such projection effects.

Second, PCXI uses the mean relation between the central 
galaxy mass and halo mass to estimate halo mass, while our halo 
masses are estimated from our halo-based group finder. 
Given that the central galaxy mass increases only slowly with 
halo mass at $M_{200}>10^{13}{\rm M}_\odot$ 
\citep{YangMoBosch03}, and the relation has significant 
amounts of scatter \citep[e.g.][]{YangMoBosch08}, binning 
based on central galaxy mass may mix halos of very different 
masses.

\citet{greco15} used a LBG sample similar to that used by PCXI,
together with the {\it Planck} temperature aperture photometries,
instead of the matched filter, to extract tSZ signals associated 
with the LBGs. They found that their results are consistent 
with the UPP model within the uncertainties of the data.
It is unclear if the difference between their results and ours 
is produced by the different mass proxies used to bin the data
or by the different methods used to extract the tSZ signals. 
\citet{vikram17} examined the cross-correlation between groups in 
the catalog of \citet{yang07} and the {\it Planck} $y$-map, and found that 
two-halo terms dominate the signals around halos of 
$M_{200}\leq 10^{13-13.5}\,h^{-1}{\rm M_\odot}$. This is in 
qualitative agreement with our finding that the stacked 
signals for low-mass groups are dominated by projection effects.

We also compare our results with results from two hydrodynamic simulations. 
The first is that presented in \citet{lebrun15}, who 
used the cosmo-OWLS suite of cosmological simulations
\citep{lebrun14}, an extension of the OverWhelmingly Large Simulations 
\citep[OWLS;][]{schaye10}, to model the tSZ effects.  
The simulation has a box size of $400h^{-1}$ Mpc on a side, and 
assumes cosmological parameters either from the {\it WMAP}7 or the 
{\it Planck}. Their fiducial runs include both stellar and AGN feedbacks. 
In Figs.~\ref{fig_Y500} and \ref{fig_gas_frac}, the predictions 
of two of their models are plotted as the two solid curves. 
The upper curves correspond to their AGN feedback model AGN8.0, which  
assumes that accreting black holes heat their surrounding gas 
to a temperature $\Delta T_{\rm heat}=10^8\,{\rm K}$, while the lower 
curves are for their AGN8.5, which assumes 
$\Delta T_{\rm heat}=3\times 10^8\,{\rm K}$. Clearly,  
our results are in a good agreement with their results, 
particularly from that of the AGN8.5 run. 

\citet{vdV16} used a suite of cosmological zoom-in simulations from 
the Feedback In Realistic Environments \citep[FIRE;][]{hopkins14, fg15, feldmann16} 
project, to study the tSZ effects around halos 
with $M_{500}=10^{10}$--$10^{13}\,{\rm M_\odot}$. Sixteen and thirty six  
zoom-in simulations were run to $z=0$ and $z\sim 2$, respectively. 
In Figs.~\ref{fig_Y500} and \ref{fig_gas_frac}, we use 
straight lines to roughly represent their low-$z$ results. 
Here the universal profile of A10 is used to convert the
predictions, which are integrated quantities 
within projected radius, to quantities within spheres
needed in the comparison. It is seen that the predicted 
tSZ signals are much stronger than both our results and 
the simulations of \citet{lebrun15}. We note, however, that 
the simulations used by \citet{vdV16} do not include  
AGN feedback, which may be important for the halo mass range
concerned here.

%%%%%%% SECTION 5
%% SHL revised ##
%% HJM revised
\section[summary]{Summary and conclusion}
\label{sec_sum}

In this paper, we use the measurements of the thermal Sunyaev-Zel'dovich 
(tSZ) effect 
from the {\it Planck} NILC all-sky Compton parameter map, together with the group 
catalogs of \citet{lim17} to investigate the hot gas contents of galaxy groups.
The catalogs contain a large number of uniformly selected groups with 
reliable halo mass estimates, which allows us to bin groups of similar halo  
masses to investigate the dependence of the tSZ effect on halo mass
over a large mass range. We adopt the matched filter 
approach \citep{haehnelt96, herranz02, melin05, melin06},
which optimizes the signal-to-noise ratio by imposing prior knowledges of 
the expected signals, to extract the tSZ signals produced by galaxy groups from the 
map. We jointly match the filters to all groups to minimize projection effects.

We test the robustness of our method by retaining or eliminating
pixels covered by local galaxy systems, by truncating the matched filters 
at different radii, by shifting the filters for low-mass groups, and 
by adding artificial signals to the observational map. We find that 
our method performs well in these tests. We also found that the background 
fluctuations around low-mass systems are significantly affected by projections of massive 
halos. Such a projection effect can lead to overestimation of the tSZ 
signals associated with low-mass groups if filters are not matched 
simultaneously to all groups.

We find that the integrated $y$-parameter and the hot gas
content it implies are consistent with the predictions of the UPP model 
only for massive groups with masses above $10^{14}\,{\rm M}_\odot$, but much 
lower, by a factor of $\sim 10$, than the model prediction for low-mass groups. 
Our results are in conflict with the findings from some previous studies 
\citep[e.g.][]{pcxi, greco15}, which reported that their data are in agreement 
with the predictions of UPP model. The disagreement 
likely comes from the different treatments of projection effects and 
the different halo mass models used in these studies.   
The halo mass dependence we find is in good agreement with the predictions 
of a set of hydro simulations presented in \citet{lebrun15} that include strong AGN 
feedback, but the simulations of \citet{vdV16}, which include only supernova 
feedback, over-predict the hot gas contents in galaxy groups by a 
factor of 5 to 10. 

 Since the integrated $y$-parameter is a measure of the thermal energy 
content of the hot halo gas, our results indicate that this energy 
content in low-mass groups is much lower than that expected from 
the universal baryon fraction in a hot halo at the virial temperature.
This has important implications for galaxy formation and evolution.
Since the total baryon fraction of stars and cold gas in galaxy 
groups and clusters is found to be well below the universal baryon 
fraction \citep[e.g.][]{FukugitaPeebles04}, it has been speculated 
that the missing baryons may be in hot defused halos. However,   
if the low energy content found here is due to a low gas content in the 
hot phase, then hot gas halos cannot account for the missing baryons.    
Alternatively, baryons originally associated with galaxy groups 
may be heated and ejected by some processes. The agreement of our 
results with the predictions of the simulation results of 
\citet{lebrun15} suggests that strong AGN feedback may be able 
to provide such a process and to accommodate the observational results. 
Yet another possibility is that a large fraction of baryons may be in phases 
with temperatures much lower than the virial temperatures of the 
groups. In this case, the low thermal energy contents observed 
in low-mass halos are produced by the low gas temperature rather 
than by a reduced amount of gas. To distinguish the different 
possibilities, it is crucial to estimate the total mass in 
the warm-hot phase, so as to obtain a complete inventory of 
the baryons in low-mass halos. This can be done either through 
quasar absorption studies, or by investigating the 
kinetic SZ effect of galaxy groups, which depends on 
the electron density but not the temperature of the halo gas.

\acknowledgments
We thank Eiichiro Komatsu for helpful discussions. 
HJM acknowledges the support from NSF AST-1517528. 
This work is also supported by the China 973 Program (No. 2015CB857002) and the 
national science foundation of China (grant Nos. 11233005, 11621303,
11522324, 11421303, 11503065, 11673015), and South Africa National Research Foundation (grant no.105925).

\end{document}